\documentclass[prl,twocolumn,nowpacs,floatfix,amsmath,amssymb,superscriptaddress]{revtex4}
\usepackage{bibunits}
\usepackage{latexsym}
\usepackage{graphicx}
\usepackage{tabularx}
\usepackage{times}
\usepackage{color}
\usepackage{amsmath}
\usepackage{dcolumn}
\usepackage{latexsym,amsmath,amssymb,bm,euscript}
\newcommand{\abs}[1]{\vert #1 \vert}

\begin{document}

\title{Exotic circuit elements from zero-modes in hybrid superconductor/quantum
Hall systems}

\author{David J. Clarke}
\affiliation{Department of Physics, California Institute of Technology,
Pasadena, CA 91125, USA}
\author{Jason Alicea}
\affiliation{Department of Physics, California Institute of Technology,
Pasadena, CA 91125, USA}
\author{Kirill Shtengel}
\affiliation{Department of Physics and Astronomy, University of California,
Riverside, CA 92521, USA}

\begin{abstract}
{Heterostructures formed by quantum Hall systems and superconductors have
recently been shown to support widely coveted Majorana fermion zero-modes
and still more exotic `parafermionic' generalizations.  Here we establish
that probing such zero-modes using quantum Hall edge states yields
\emph{non-local} transport signatures that pave the way towards a variety
of novel circuit elements.  In particular, we demonstrate quite generally
that at low energies the zero-modes convert chirally moving quasiparticles
into oppositely charged quasiholes propagating in the same direction---that
is, they swap the sign of the chiral edge currents. One may then construct
new and potentially useful circuit elements using this `perfect Andreev
conversion' process, including superconducting current and voltage mirrors
as well as transistors for fractional charge currents. Characterization of
these circuit elements should provide striking evidence of the zero-mode
physics.}
\end{abstract}
\maketitle
\defaultbibliographystyle{apsrev}
\defaultbibliography{kirref}
\begin{bibunit}

\textbf{\emph{Introduction.}}\ Non-Abelian anyons provide a fascinating
illustration of Anderson's `more is different' paradigm~\cite{Anderson1972}.
These quasiparticles, which emerge from interacting collections of ordinary
bosons and fermions, produce a ground-state degeneracy that scales
exponentially with the number of anyons present in the host system.
Moreover, braiding the anyons around one another noncommutatively rotates the
system's quantum state within this degenerate manifold.  These remarkable
properties have led to great interest in non-Abelian anyons for use in
fault-tolerant quantum information processing devices~\cite{Nayak2008}.  Our
goal here is to propose another application of non-Abelian anyons, namely the
construction of unusual circuit elements such as transistors for fractional
charge, current/voltage mirrors, and flux-based capacitors.

\begin{figure}
\includegraphics[width = \columnwidth]{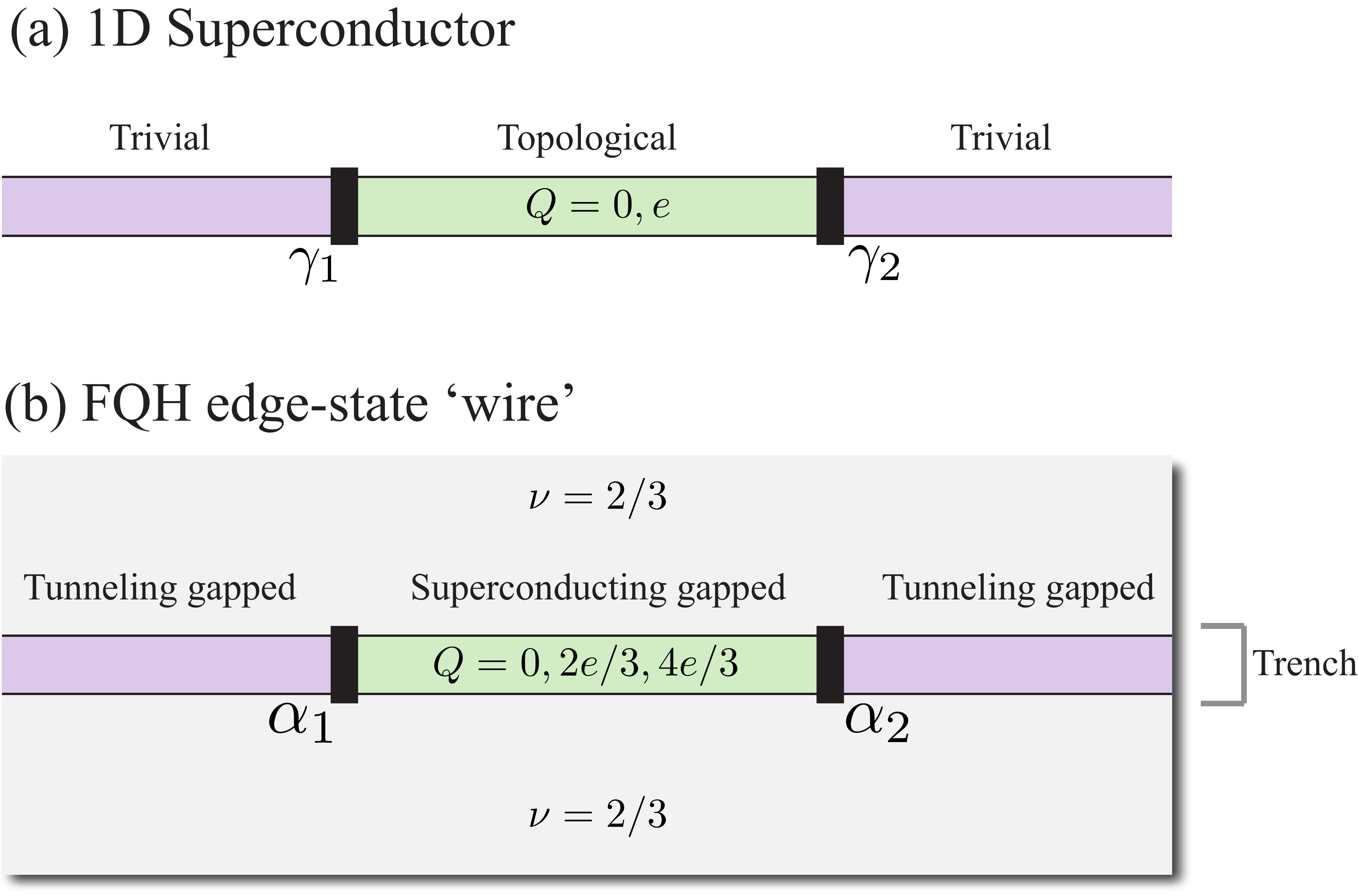}
\caption{(a) Majorana zero-modes $\gamma_j$ bind to domain walls separating topological
and trivial segments of a 1D superconducting wire.
(b) `Parafermion' zero-modes $\alpha_j$, which represent $\mathbb{Z}_3$
generalized Majorana modes, localize
between segments of a trench in a $\nu=2/3$ quantum Hall state gapped by
tunneling and Cooper pairing.  In both cases the zero-modes encode a degeneracy
among states with different charge $Q$ (mod $2e$) on the central region,
as labeled in the figure.}
\label{FigZeroModes}
\end{figure}

As a primer, let us first consider a one-dimensional (1D) topological
superconductor \cite{Kitaev2001}, obtained when an odd-channel wire acquires
a bulk pairing gap. Suppose that a 1D superconductor breaks up into
alternating topological and trivial domains as in Fig.~\ref{FigZeroModes}(a)
(\emph{e.g.}, by varying the number of channels spatially along the wire).
Here the endpoints of the topological regions realize `Ising' non-Abelian
anyons, which bind Majorana zero-modes that encode a two-fold ground-state
degeneracy per topological segment.  Physically, the degeneracy reflects the
fact that each topological domain can switch its fermion parity without
affecting the energy density---contrary to conventional superconductors.
Numerous sources of 1D topological superconductivity have been proposed,
which involve coupling a bulk superconductor to systems such as
two-dimensional (2D) topological insulator edges~\cite{Fu2008},
spin-orbit-coupled nanowires~\cite{Lutchyn2010a,Oreg2010}, magnetic-atom
chains~\cite{Choy2011,Nadj-Perge2013}, or counterpropagating sets of integer
quantum Hall edge modes~\cite{Clarke2013a, Lindner2012} (for reviews see
Refs.\ \cite{Beenakker2013a,Alicea2012a,Leijnse2012,Stanescu2013a}).

Among these platforms, the integer quantum Hall architecture most naturally
generalizes to fractionalized setups that harbor richer phenomena stemming from
the interplay between superconductivity and strong correlations. Consider, for
instance, a `wire' synthesized from counterpropagating fractional quantum Hall
edge states separated by a narrow trench [see Fig.~\ref{FigZeroModes}(b) for an
example at filling $\nu = 2/3$].  This `wire' can acquire a gap either through
electron backscattering across the trench or via Cooper pairing. Because the
edge states support fractionally charged excitations, the ends of
pairing-gapped regions correspond to exotic non-Abelian anyons binding
generalizations of Majorana
zero-modes~\cite{Clarke2013a,Lindner2012,Cheng2012}.  These
\emph{parafermionic} zero-modes~\cite{Fendley2012} encode a larger ground-state
degeneracy compared to the usual Majorana case, since here each
superconducting-gapped region can acquire \emph{fractional} charge without
changing its energy density.  Similar effects may arise in other fractionalized
setups~\cite{Barkeshli2012a,Vaezi2013,Oreg2013,Klinovaja2014a,Klinovaja2014b},
including quantum Hall bilayers in which interlayer tunneling plays the role of
Cooper pairing~\cite{Barkeshli2013a,Barkeshli2013b}.

\begin{figure}
\includegraphics[width = 3in]{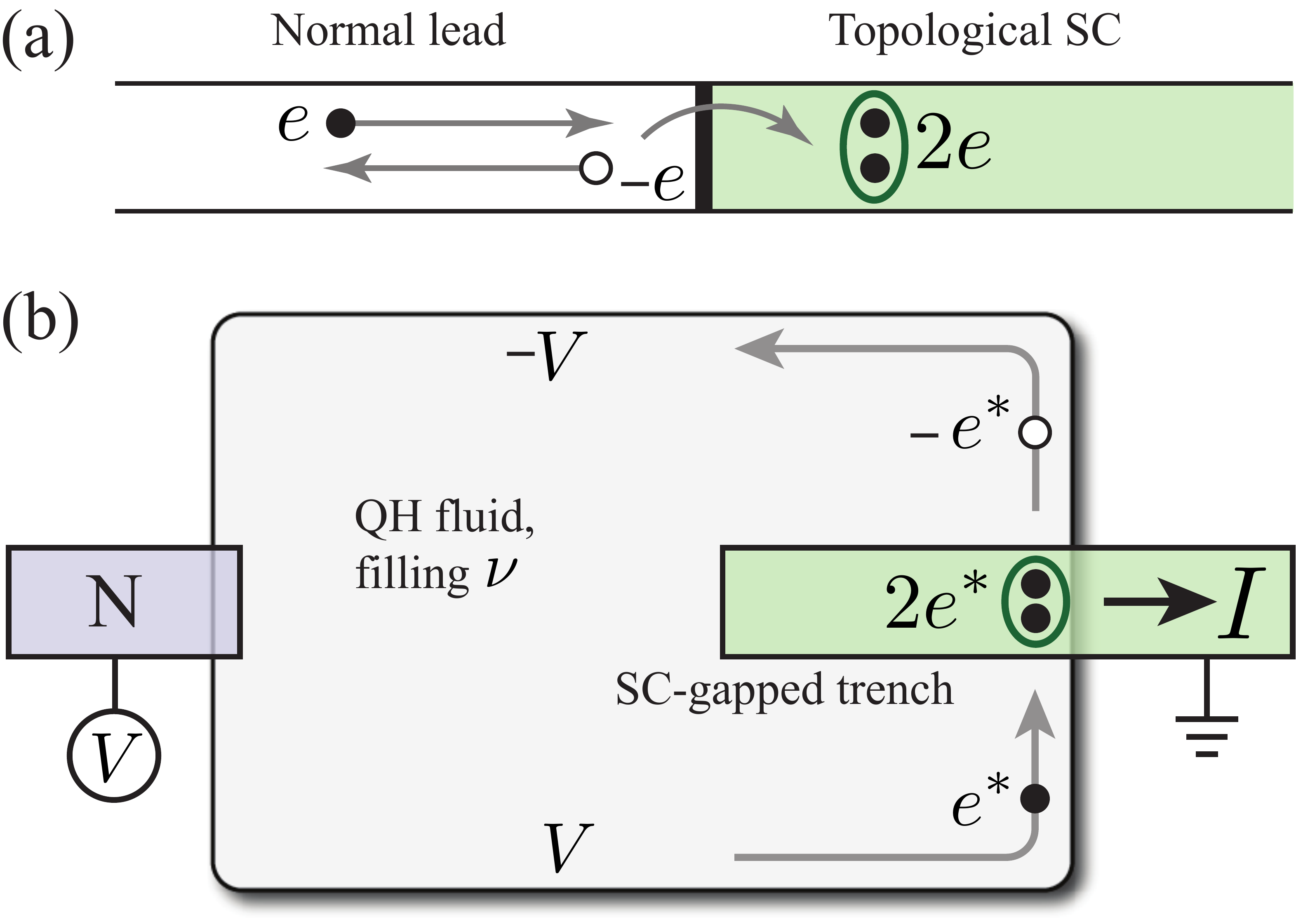}
\caption{(a) Electrons injected from a normal lead towards a 1D topological
superconductor perfectly Andreev reflect into holes due to coupling with
a Majorana zero-mode at the junction.
(b) The zero-mode in a pairing-gapped trench of a quantum Hall
fluid similarly mediates `perfect Andreev conversion', transforming incoming
quasiparticles from the edge into outgoing quasiholes.
In either case, the superconductor absorbs the excess charge.}
\label{FigAndreev}
\end{figure}

In this paper we predict novel \emph{non-local} transport signatures of
Majorana and parafermionic zero-modes in quantum Hall/superconductor hybrids
that, in turn, provide a foundation for the unusual circuit elements
mentioned above. The experiments we propose relate closely to the `zero-bias
anomaly'  arising when a single-channel normal lead probes a Majorana
zero-mode in a 1D topological
superconductor~\cite{Sengupta2001,Bolech2007,Law2009,Flensberg2010,
Wimmer2011,Fidkowski2012,Lin2012,Affleck2013,Lutchyn2013}.
In such a setup---sketched in Fig.~\ref{FigAndreev}(a)---the Majorana mode is
predicted to mediate perfect Andreev reflection as temperature $T$ and bias
voltage $V$ approach zero.  That is, in this asymptotic limit an incoming
electron from the normal lead reflects off of the topological superconductor
as a hole with unit probability, yielding a quantized zero-bias conductance
of $2e^2/h$. Importantly, this is twice the conductance of an ideal
single-channel wire, with the factor of two arising because of the added
contribution of the reflected hole.

We show that quantum Hall/superconductor hybrids yield an interesting variation
on this transport anomaly, in particular when the native edge states serve as a
lead that probes zero-modes generated in such setups. The basic transport
architecture appears in Fig.~\ref{FigAndreev}(b) and contains two new features
compared to the 1D topological superconductor problem. First, in the fractional
quantum Hall case the (parafermionic) zero-mode at the outer trench edge
mediates perfect conversion of incoming quasiparticles carrying
\emph{fractional} charge $e^*$ into $-e^*$ quasiholes as $T,V \rightarrow 0$,
thereby transmitting charge $2e^*$ into the superconductor. Such events are
possible since the pairing-gapped trench can resonantly absorb fractional
charge due to the ground-state degeneracy. (Ref.~\cite{Barkeshli2013b} briefly
explored an analogous transport phenomenon in a bilayer setup.)  Second---and
more importantly---if the (ungapped) quantum Hall edge used as a lead supports
purely chiral charge excitations, the outgoing quasihole continues in the same
direction around the edge as Fig.~\ref{FigAndreev}(b) illustrates. We refer to
this process as \emph{Andreev conversion} to distinguish it from standard
Andreev reflection in which the hole retraces the incoming particle's path
\footnote{Andreev conversion is essentially a chiral analogue of crossed
Andreev reflection~\cite{Nilsson2008,Herrmann2010,Das2012b}.}. The
superconductor and quantum Hall edges thus form a trijunction in which the
current and voltage in each leg exhibits a strong dynamical constraint. When
multiple superconducting trenches are immersed in the same quantum Hall fluid,
this dynamical constraint underlies various non-local transport anomalies.  In
what follows we theoretically establish the perfect Andreev conversion noted
above under rather general circumstances and then discuss several novel circuit
elements that follow.

\textbf{\emph{Perfect Andreev conversion}.}\  One can access the transport
phenomena we describe using many different quantum Hall phases.  The
conceptually simplest corresponds to a $\nu = 1$ integer quantum Hall state,
though coupling to superconductivity in this case appears non-trivial due to
spin polarization.  Alternatively, the $\nu = 1$ edge mode can arise from the
magnetically gapped surface of a 3D topological insulator
\cite{Fu2009b,Akhmerov2009a}.  In this realization one can utilize ordinary
$s$-wave superconductors; moreover, orbital magnetic fields are not required.
Quantum Hall phases with multiple edge channels also suffice.  We will simply
require that the edge supports a single, chiral charge mode that at low
energies decouples from all other neutral modes.  This decoupling occurs,
\emph{e.g.}, for hierarchical states at $\nu = n/(2np+1)$ ($n$ and $p$ are
integers) due to disorder as shown by Kane and Fisher \cite{Kane1995}.  Among
this series the spin-unpolarized $\nu = 2/3$ state is particularly
advantageous, in part because here too one can induce pairing via an $s$-wave
superconductor.  Another virtue is that the instabilities leading to the
zero-modes occur at weak coupling for a range of realistic parameters.  We
expound on this important technical point in the Supplementary Information,
which explores the $\nu = 2/3$ case in greater detail.

Suppose that we etch a trench into a quantum Hall system and then fill the
void with a superconductor as shown in Fig.~\ref{FigAndreev}(b).  Provided
the trench width does not exceed the superconductor's coherence length, the
proximity effect can then gap out the adjacent edge modes through Cooper
pairing.  As noted earlier one thereby creates zero-modes at the ends of the
trench that encode a ground-state degeneracy for the intervening
pairing-gapped region.    We will now explore transport resulting when the
gapless chiral charge mode impinging on the trench hybridizes with one of
these zero-modes.

The key physical mechanism is that at low energies the gapped trench imposes
certain boundary conditions that relate incoming and outgoing quasiparticles
from the adjacent gapless edge.  Consider first the limit in which the
gapless edge state completely decouples from the superconducting trench.
Concretely, one could envision adding near the boundary a small
tunneling-gapped region as in Fig.~\ref{FigZeroModes}(b) to block coupling to
the zero-mode.  In this case incoming quasiparticles continue along the edge
uninterrupted by the trench, and the action governing the charge mode of
interest is simply that of an unperturbed edge \cite{WenBook}:
\begin{equation}
  S_{\mathrm{charge}}=\frac{-1}{4\pi\nu}\int\!\mathrm{d}x\,\mathrm{d}t\, \partial_x\phi
  \left(\partial_t+v\partial_x\right)\phi,
  \label{Lcharge}
\end{equation}
with $\phi$ a field that determines the total edge charge density through
$\rho = e\partial_x \phi/(2\pi)$.  Commutation relations implicit in the
action imply that $e^{i\phi}$ is an operator that adds charge $e^* = \nu e$.
Throughout we assume that this charge mode decouples from all neutral
modes---should any exist.  Equation~(\ref{Lcharge}) describes a fixed point
for the edge at which particles undergo `perfect normal transmission' upon
hitting the trench.

Next we incorporate weak coupling to the zero-mode nearest to the edge, which
allows charge $e^*$ to pass between the gapless edge and the pairing-gapped
region.  Such processes perturb the above fixed-point action with a term
\begin{equation}
  \delta S_{\rm zero-mode} = \Gamma\int \mathrm{d}t\left[e^{i\Phi}
  e^{-i\phi(x = x_0,t)} + \text{H.c.}\right],
  \label{deltaL}
\end{equation}
where $x_0$ is the position where the trench and gapless charge mode
intersect, $\Gamma$ denotes the bare coupling strength, and $e^{i\Phi}$ is an
operator that cycles the charge on the pairing-gapped region of the trench by
$e^*$ (mod $2e$).   Under renormalization $\Gamma$ flows according to
$\partial_\ell \Gamma = (\Delta - 1)\Gamma$, where $\ell$ is a logarithmic
rescaling factor and $\Delta = \nu/2$ is the scaling dimension of the
tunneling operator in Eq.\ (\ref{deltaL}).  For $\nu < 2$---to which we
specialize hereafter---hybridization with the zero-mode thus constitutes a
relevant perturbation that destabilizes the perfect normal transmission fixed
point.

The system then flows to a different fixed point at which coupling to the
zero-mode imposes nontrivial boundary conditions on the gapless charge mode
at $x = x_0$.  This boundary condition can be expressed as
\begin{equation}
  \phi_\mathrm{out}=2\Phi-\phi_\mathrm{in}
  \label{Eqboundary}
\end{equation}
with $\phi_{\rm out/in} \equiv \phi(x = x_0 \pm 0^+)$ denoting gapless
charge-sector fields evaluated just above and below the trench. Equation
(\ref{Eqboundary}) causes incoming charge-$e^*$ quasiparticles ($e^{i\phi}$)
\emph{added relative to the superconductor's potential} to continue as
outgoing $-e^*$ quasiholes ($e^{-i\phi}$), with the pairing-gapped trench
absorbing the deficit charge $2e^*$ ($e^{2i\Phi}$).  This is precisely the
Andreev conversion process described earlier. There is, however, a competing
effect, whereby electrons tunnel directly past the trench.  We now address
the relevance of these processes at the `perfect Andreev conversion' fixed
point defined by the boundary condition in Eq.\ (\ref{Eqboundary}).

Any perturbation that tunnels an electron \emph{past} the trench takes the form
\begin{eqnarray}
  \delta S_{\rm tunneling} &=& \lambda\int \mathrm{d}t
  \left[e^{i\phi_\mathrm{in}/\nu-i\phi_\mathrm{out}/\nu}\mathcal{O}_n+\text{H.c.}\right]
  \nonumber \\
  &\rightarrow & \lambda\int \mathrm{d}t
  \left[e^{2i(\phi_\mathrm{in}-\Phi)/\nu}\mathcal{O}_n+\text{H.c.}\right],
  \label{Stunneling}
\end{eqnarray}
where $\lambda$ is the coupling strength and $\mathcal{O}_n$ contains
possible neutral parts of the electron tunneling operator (\emph{e.g.},
operators that transfer spin).  In the second line we used Eq.\
(\ref{Eqboundary}) to eliminate $\phi_{\rm out}$. The scaling dimension of
the above electron tunneling term reads $\Delta_e=(2/\nu)+\Delta_{n}$, with
$\Delta_{n}$ the scaling dimension of $\mathcal{O}_n$.  Consequently, any
operator that transfers electron charge across the trench has
$\Delta_e>2/\nu$ and is hence irrelevant for $\nu<2$---implying stability of
the perfect Andreev conversion fixed point in the regime of interest
\footnote{Actually, the neutral sector exhibits trivial boundary conditions
at the trench which allows one to eliminate $\mathcal{O}_n$ as well in Eq.\
(\ref{Stunneling}).  Hence $\Delta_n = 0$, implying instability of the
Andreev fixed point when $\nu >2$ (where the normal fixed point is stable).}.
Numerous interesting consequences for transport follow from this general
result.

\textbf{\emph{Doubled Hall voltage.}}\ Consider again the setup of
Fig.~\ref{FigAndreev}(b), where a normal electrode at potential $V$ injects
charge along the lower quantum Hall edge towards a grounded superconducting
trench. We temporarily assume negligible contact resistances; the lower edge
is then also at potential $V$.  For `small' $V$ (quantified below), injected
charge undergoes perfect Andreev conversion at the trench.  Thus the
conductance $G$ \emph{doubles} compared to a standard two-terminal
measurement---\emph{i.e.}, $G = 2 \sigma_\text{H}$, where $\sigma_\text{H} =
\nu e^2/h$ is the Hall conductance.  For $\nu = 1$ the physics essentially
maps to the 1D topological superconductor case described earlier, modulo the
spatial separation of incoming and outgoing modes.

The potential on the upper quantum Hall edge, where the outgoing
Andreev-converted carriers flow, follows from current conservation.  More
precisely, since the same current $I = G V$ passing through the
superconductor traverses the quantum Hall fluid, the current must
additionally satisfy $I = \sigma_\text{H} V_H$ ($V_H$ is the Hall voltage).
Consistency requires that the Hall voltage, like the conductance, is also
doubled compared to the case where both source and drain are normal
electrodes.  That is, the potential on the upper edge is $-V$, opposite that
of the lower edge.  This implies that \emph{the Andreev-converted carriers do
not equilibrate with the superconductor}; rather equilibration occurs at the
normal electrode in Fig.~\ref{FigAndreev}(b).

Consider next the more general situation where the potential for the
superconductor is $V_\text{SC}$ and that of the incoming/outgoing quantum
Hall edges is $V_{{\rm in/out}}$.  Due to the doubled conductance the current
flowing out of the superconductor is $I =
2\sigma_\text{H}(V_{\mathrm{in}}-V_\text{SC})$, while the (same) current
crossing the Hall fluid obeys $I = \sigma_\text{H}(V_{\rm in}-V_{\rm out})$.
It follows that
\begin{equation}\label{eqV}
V_{\mathrm{out}}=2V_\text{SC}-V_{\mathrm{in}},
\end{equation}
which we will frequently employ in the forthcoming discussion.  These results
apply when temperature and the superconductor/quantum Hall voltage
differences are $(i)$ small compared to the zero-mode hybridization energy
$\Gamma$ [recall Eq.\ (\ref{deltaL})] and $(ii)$ sufficiently small that the
irrelevant electron tunneling terms [Eq.\ (\ref{Stunneling})] remain
inoperative; this is the regime where perfect Andreev conversion holds.  Note
also that imperfect superconducting contact will decrease the voltage drop
compared to Eq.\ (\ref{eqV}).  In general,
$V_{\mathrm{out}}=V_{\mathrm{in}}+\eta(V_\text{SC}-V_{\mathrm{in}})$,
where the quality factor $\eta$ ranges between 0 (no contact) and 2 (ideal
superconducting contact).  We stress that if both electrodes were normal,
$\eta$ could never exceed $1$.  Unless otherwise stated, we will assume ideal
superconducting contacts in the devices described below to emphasize their
unusual behavior.

\begin{figure}
\includegraphics[width = \columnwidth]{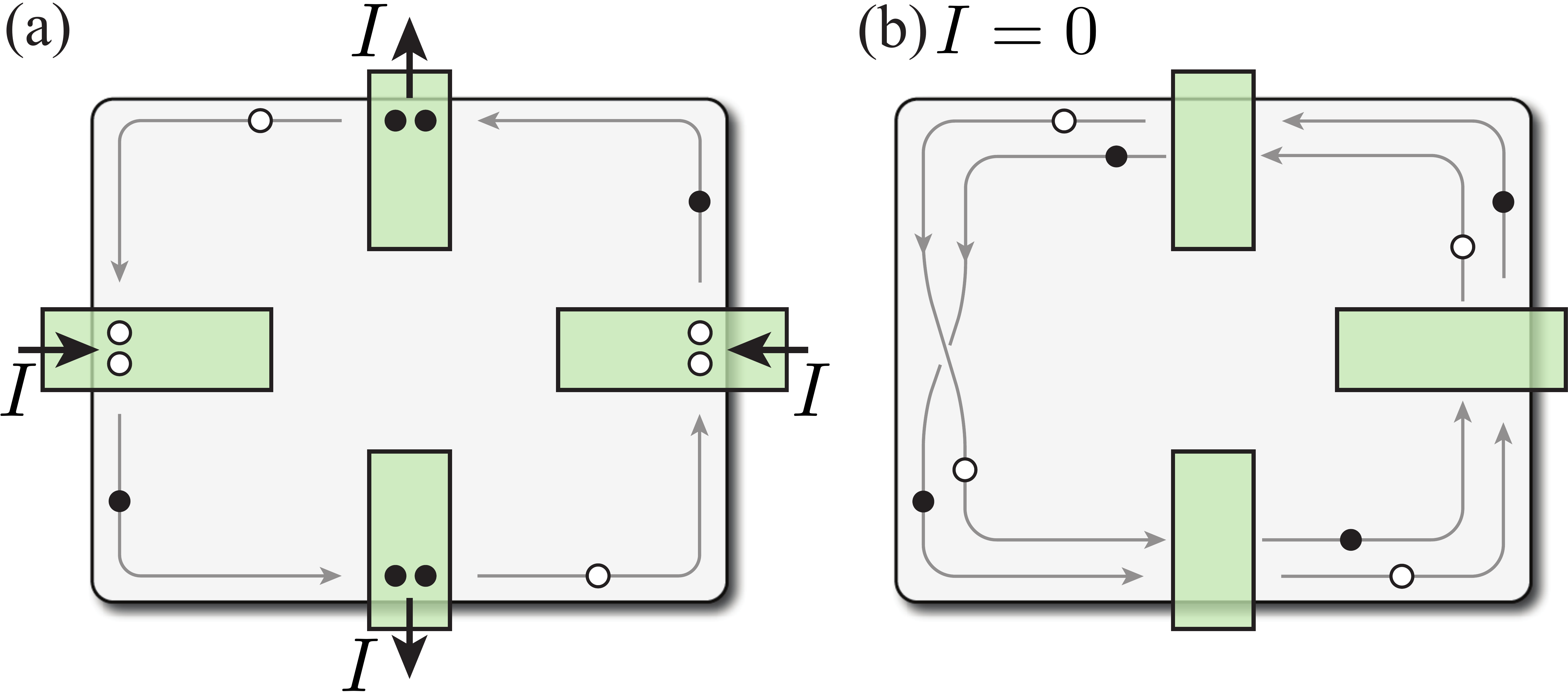}
\caption{(a) Quantum Hall fluid with four pairing-gapped trenches generated by
equipotential superconductors.
If each superconductor induces perfect Andreev conversion, the same current
$I$ must flow through all four superconductors, with relative orientation
indicated by the large arrows.  (b) If an odd number of equipotential
superconductors enter the quantum Hall fluid, the system can no longer carry
dc current, as the charge added to the superconductors due to Andreev conversion
time-averages to zero.  }
\label{Fig0Transformer}
\end{figure}

\textbf{\emph{Transistor for fractional charge.}}~Figure~\ref{Fig0Transformer}
illustrates a simple example of non-local effects resulting when multiple
superconducting trenches appear in the quantum Hall system.  Here, all
superconductors are set to the same potential. Unlike in the Josephson effect,
their relative phases are inconsequential -- we assume that any notion of phase
coherence is lost along the edges connecting the leads. The charge, however, is
conserved in such transport and, as a consequence, in
Fig.~\ref{Fig0Transformer}(a)---where the number of superconductors is
even---current can only flow from one superconductor to another if it flows in
all four simultaneously, with relative orientations specified in the figure.
Sending current from the left to bottom superconductors, for instance,
necessarily yields the same current flow from right to top, a behavior
reminiscent of the non-local transport mediated by excitons in bilayer quantum
Hall systems~\cite{Finck2011}. The restricted current arises because $(i)$
quasiparticles undergo perfect Andreev conversion at each trench and $(ii)$ a
given superconductor receives the \emph{same} charge under subsequent `round
trips' along the edge. By contrast, with an odd number of equipotential
superconductors, as in Fig.~\ref{Fig0Transformer}(b), (direct) current can
simply not flow.  This becomes evident upon tracing the path of a single
quasiparticle around the edge.  On the first round trip, the quasiparticle
deposits charge on each superconductor due to Andreev conversion, but on the
next pass \emph{removes} these same charges because the number of
superconductors is odd. Figure~\ref{Fig0Transformer} illustrates the relevant
quasiparticle processes in both cases.  One can utilize this even/odd effect to
create a transistor for fractional charge by using gates to controllably
isolate one of the superconductors from the rest of the system.

\begin{figure}
\includegraphics[width = \columnwidth]{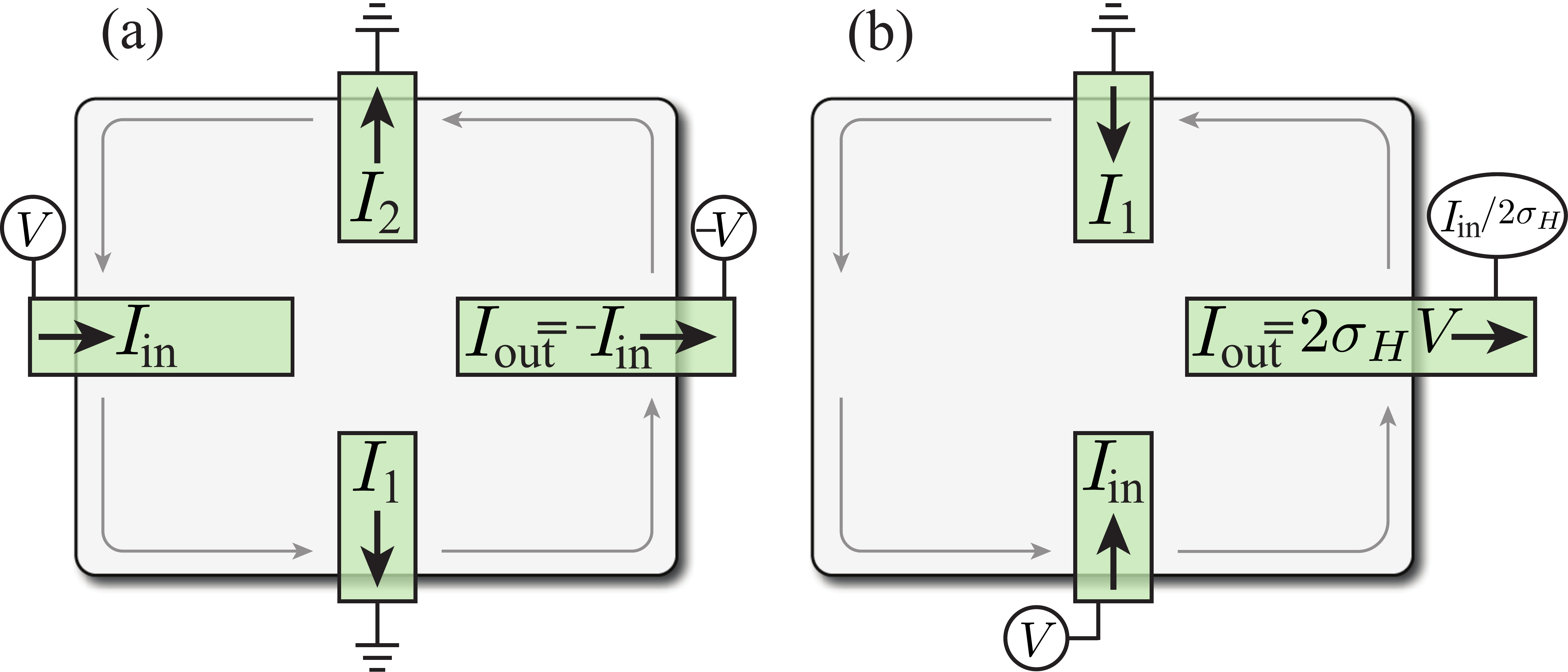}
\caption{(a) Current/voltage mirror and (b) current/voltage swap circuit
elements generated from perfect Andreev conversion.}
\label{FigMirror}
\end{figure}

\textbf{\emph{Voltage/Current Mirror.}}\ Suppose we have a device with four
superconducting trenches [as in Fig.~\ref{Fig0Transformer}(a)] but now allow
their potentials to vary.  The constraints imposed by Eq.~(\ref{eqV}),
together with the doubled conductance relating voltage to current, leave the
system with only three remaining degrees of freedom.
The first corresponds to current flow in the pattern shown by
Fig.~\ref{Fig0Transformer}(a). A second degree of freedom appears in
Fig.~\ref{FigMirror}(a), where we ground the upper and lower superconductors
while raising the voltage of the left superconductor (the control). The
voltage of the remaining superconductor (the output) necessarily goes down,
mirroring the change.  As in Fig.~\ref{Fig0Transformer}(a), the current
carried away by the output is opposite that flowing into the control; excess
current flows to ground via the top and bottom superconductors. We are left,
then, with a device that \emph{reverses} both the voltage and the current
flow from input to output. One can access the final degree of freedom by
changing the relative voltage of the two grounded superconductors. The
response to such a change may be interpreted as a superposition of
Fig.~\ref{FigMirror}(a) with the same setup rotated by $90^\circ$.

\textbf{\emph{Voltage/Current Swap.}}\ In a system with three ideal
equipotential superconducting trenches, we have already shown that no current
can flow [recall Fig.~\ref{Fig0Transformer}(b)]. However, suppose that---as
in Fig.\ \ref{FigMirror}(b)---we ground only the top superconductor and set
the current and voltage on the bottom (control) superconductor.  The third
superconductor functions as the output. In this case the constraints of
Eq.~(\ref{eqV}) lead to the unusual result that the two independent
information channels (for a superconductor), current and voltage, are swapped
at the output relative to the input.  For instance, if the current flowing
into the control terminal is $I_{\rm in}$ then the voltage at the output is
$I_{\rm in}/(2\sigma_\text{H})$; likewise, if the control terminal voltage is
$V$, the output current is $I_{\rm out} = 2\sigma_\text{H} V$.  (Excess
current again flows to ground.)

\textbf{\emph{Superconducting Flux-based Capacitor.}}\  Consider next
Fig.~\ref{FigMirror}(b) in the limit where the right superconductor is also
held to ground.  In this case the input current $I_{\rm in}$ must clearly
vanish. Nevertheless, the voltage $V$ on the bottom control superconductor
induces a current $2\sigma_\text{H} V$ flowing from the top to the right
superconducting contacts, \emph{even though the latter remain at the same
potential}.  Note that this is consistent with Fig.~\ref{Fig0Transformer}(b)
since in the limit $V = 0$ all currents vanish.  This effect persists for
larger odd numbers of superconductors as well, in which case current of
magnitude $2\sigma_\text{H} V$ flows in all grounded terminals, with a sign
alternating from lead to lead.

\begin{figure}
\includegraphics[width = 3in]{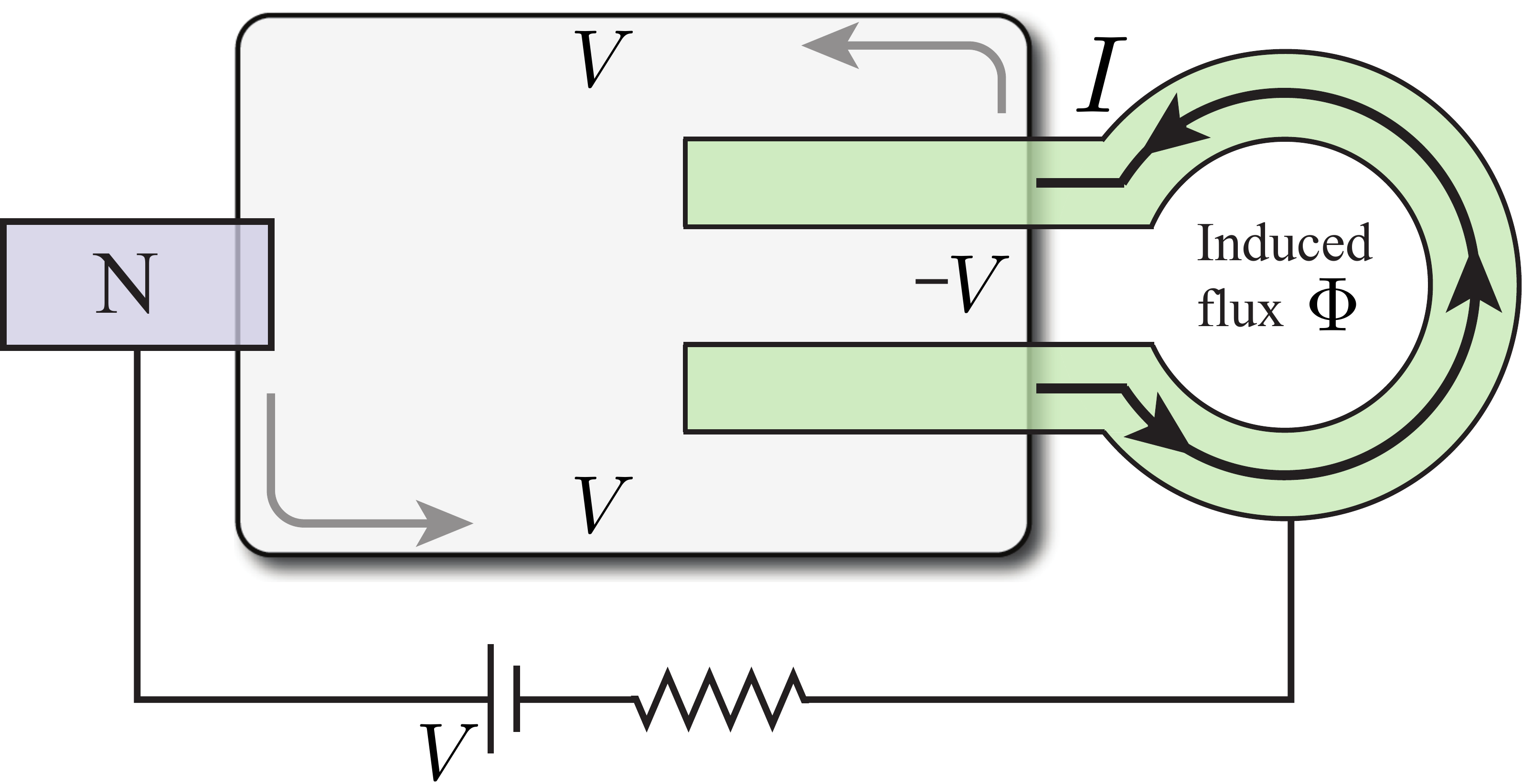}
\caption{A voltage $V$ applied to the normal electrode on the left generates
current in the superconducting loop on the right.  The induced flux is
proportional to $V$---hence the circuit acts as a `flux capacitor'.
Unlike a standard inductor current need not flow through the battery.  }
\label{FigFluxCapacitor}
\end{figure}

Finally, let us examine the circuit in Fig.~\ref{FigFluxCapacitor}, where we
use analogous physics to create a `flux capacitor' in which a voltage stores
magnetic flux. Here a pair of superconducting trenches connect into a loop on
the right side, while a normal electrode on the left is held at potential $V$
relative to the superconductors by a battery.  Once again using
Eq.~(\ref{eqV}) to determine the voltages along the circuit, we find that
current $I = 2\sigma_\text{H} V$ flows around the superconducting loop,
generating flux $\Phi \equiv C_\Phi V$.  The `flux capacitance' $C_\Phi$ is
determined by the loop's inductance.  We emphasize that unlike an ordinary
inductor, no current flows through the power source or resistor at steady
state in the ideal $\eta=2$ device. Rather, the applied voltage merely sets
the level of circulating current in the superconducting loop.

\textbf{\emph{Discussion.}}\ It is worth reiterating that the non-local
transport anomalies in the devices proposed above originate from zero-modes
bound to non-Abelian defects induced by the superconductors. Characterization
of these circuit elements is therefore a natural first step in the pursuit of
topological quantum computation with such systems
\cite{Clarke2013a,Lindner2012,Cheng2012,Mong2013}.  Another tantalizing
application is the construction of low-power logical circuits.  These circuits
would have an advantage in the control of low-temperature quantum information
devices, as they would coexist within the same low-temperature environment.
Depending on the implementation, it may also be possible to produce such
logical circuits `on-chip' with quantum information implementations already
based on 2D-electron gas and/or superconducting elements.  These applications
likely extend to alternative setups as well. The topological-insulator-based
interferometers proposed in Refs.~\cite{Fu2009b,Akhmerov2009a} yield a related
mechanism for perfect Andreev conversion, and thus should also form the
backbone of nontrivial circuit elements.  Quantum Hall bilayers (\emph{without}
superconductivity) provide another promising venue.  Novel dc transformers were
proposed                   in this context a decade ago~\cite{Halperin2003},
and given that a bilayer variant of Andreev conversion is already theoretically
established \cite{Barkeshli2013b} other interesting elements are likely also
possible.

\textbf{\emph{Acknowledgements.}}\  We are indebted to J.~P.~Eisenstein,
M.~P.~A.~Fisher, C.~Nayak and A.~Stern for numerous enlightening discussions.
We also acknowledge funding from the NSF through grants DMR-1341822 (D.~J.~C.\
\& J.~A.) and DMR-0748925 (K.~S.); the Alfred P.~Sloan Foundation (J.~A.); the
DARPA QuEST program (K.~S.); and the Caltech Institute for Quantum Information
and Matter, an NSF Physics Frontiers Center with support of the Gordon and
Betty Moore Foundation.

\putbib
\end{bibunit}

\setcounter{equation}{0} \setcounter{figure}{0}
%
\makeatletter 
\def\tagform@#1{\maketag@@@{(S\ignorespaces#1\unskip\@@italiccorr)}}
\makeatother
\makeatletter \makeatletter \renewcommand{\fnum@figure}
{\figurename~S\thefigure} \makeatother

\renewcommand{\bibnumfmt}[1]{[S#1]}
\renewcommand{\citenumfont}[1]{S#1}


\begin{bibunit}

\section{Supplementary Information}
\label{sec:supp}
In this section we characterize the instabilities leading to parafermionic
zero-modes in a spin-unpolarized $\nu = 2/3$ quantum Hall system, which for
reasons elucidated below is a particularly promising platform for the
transport anomalies described in the main text.  To begin we summarize the
edge theory for this quantum Hall state using $K$-matrix
formalism~\cite{Wen1992a}. The edge modes of the unpolarized $2/3$ state are
described by two bosonic fields $\phi_{\uparrow,\downarrow}$ and an
associated $K$-matrix $\mathbf{K}=\begin{pmatrix} 1 & 2\\ 2 & 1
\end{pmatrix}$ and charge vector $\mathbf{q}=\begin{pmatrix}
1\\1\end{pmatrix}$. In this description, the edge electron density is given
by $({1}/{2\pi})\,\mathbf{q}^\intercal \partial_x\boldsymbol{\phi}$ and the
filling fraction reads {$\nu=\mathbf{q}^\intercal \mathbf{K}^{-1}
\mathbf{q}=2/3$}.  A trench etched in the quantum Hall liquid brings two
counterpropagating sets of such modes into close proximity with one another.
Disregarding tunneling and pairing terms for the moment, the Lagrangian
density for the edge modes opposite the trench is
\begin{equation}
\mathcal{L}_0= \frac{1}{4\pi}\partial_x\phi_I
\left(\mathcal{K}_{IJ}\partial_{t}\phi_J-V_{IJ}\partial_x\phi_J\right),
\label{L0}
\end{equation}
where we have four bosonic fields $\phi_I$, two for each side of the trench.
The extended $K$-matrix for this doubled system is
$\boldsymbol{\mathcal{K}}=\begin{pmatrix} \mathbf{K} & 0\\ 0&
-\mathbf{K}\end{pmatrix}$, while the matrix $\mathbf{V}$ captures the
density-density interactions both within each edge and across the trench.

For simplicity, we assume the two edges are symmetric and that the
interactions are invariant under SU(2) spin rotations.  (Note that an
equivalent SU(2) symmetry \emph{emerges} due to disorder in the polarized
$\nu = 2/3$ state~\cite{Kane1994b}; here, however, the SU(2) symmetry is
manifestly that of spin.)  This gives us the generic for m
\begin{equation}\label{Vdef}
\mathbf{V}=\begin{pmatrix} v_1 & v_2 & v_3 & v_4\\ v_2 & v_1 & v_4 & v_3\\
v_3 & v_4 & v_1 & v_2\\ v_4 & v_3 & v_2 & v_1\end{pmatrix}.
\end{equation}
Of course the SU(2) spin symmetry is not guaranteed microscopically.  In
particular, this symmetry is broken by a Zeeman field, which takes the form
$H_\text{Zeeman}=({h}/{2\pi})
\mathbf{n}_\text{t}^\intercal\partial_x\boldsymbol{\phi}$, with
$\mathbf{n}_\text{t}^\intercal=(1,-1,1,-1)$. This term has scaling dimension
1, but can be absorbed into $\mathcal{L}_0$ via a redefinition
$\boldsymbol{\phi}\rightarrow\boldsymbol{\phi}-hx\mathbf{V}^{-1}
\mathbf{n}_\text{t}$. However, this absorbtion will cause other terms
involving $\exp{(i \mathbf{n}_\text{t}^\intercal \boldsymbol{\phi})}$ to
oscillate on the length scale $\pi(v_1-v_2+v_3-v_4)/2h$.

Next we classify perturbations to Eq.\ (S\ref{L0}) that can generate
instabilities.  There are six types of gap-opening perturbations that are
marginal when $h=0$ and density--density interactions between the two edges
are absent, \emph{i.e.} $v_3=v_4=0$. Charge hopping and pairing terms each
come in two varieties due to spin degeneracy. We can divide these into
perturbations that form either spin singlets or spin triplets across the
trench. In addition, there are marginal perturbations that do not transfer
charge but involve singlet or triplet spin correlations across the trench.
These perturbations are listed in Table~\ref{table1}, along with their
scaling dimensions (expressed in terms of parameters $u$ and $v$ defined in
the caption).

\begin{table}[htbp]
\begin{tabularx}{\columnwidth}{|l|l|X|}
\hline
Process&Operators&Scaling Dimension\\
\hline
singlet hopping
& $e^{\pm\frac{1}{2}\mathbf{c}_\text{h}^\intercal\boldsymbol{\mathcal{K}}\boldsymbol{\phi}}
e^{\pm\frac{1}{2}\mathbf{n}_\text{s}^\intercal\boldsymbol{\mathcal{K}}\boldsymbol{\phi}}$
& $\frac{1}{2}e^{2u}+\frac{3}{2}e^{2v}$
\\
triplet hopping
& $e^{\pm\frac{1}{2}\mathbf{c}_\text{h}^\intercal\boldsymbol{\mathcal{K}}\boldsymbol{\phi}}
e^{\pm\frac{1}{2}\mathbf{n}_\text{t}^\intercal\boldsymbol{\mathcal{K}}\boldsymbol{\phi}}$
& $\frac{1}{2}e^{-2u}+\frac{3}{2}e^{2v}$
\\
singlet pairing
&$e^{\pm\frac{1}{2}\mathbf{c}_\text{p}^\intercal\boldsymbol{\mathcal{K}}\boldsymbol{\phi}}
e^{\pm\frac{1}{2}\mathbf{n}_\text{s}^\intercal\boldsymbol{\mathcal{K}}\boldsymbol{\phi}}$
& $\frac{1}{2}e^{2u}+\frac{3}{2}e^{-2v}$
\\
triplet pairing
&$e^{\pm\frac{1}{2}\mathbf{c}_\text{p}^\intercal\boldsymbol{\mathcal{K}}\boldsymbol{\phi}}
e^{\pm\frac{1}{2}\mathbf{n}_\text{t}^\intercal\boldsymbol{\mathcal{K}}\boldsymbol{\phi}}$
& $\frac{1}{2}e^{-2u}+\frac{3}{2}e^{-2v}$\\
neutral singlet&$e^{\pm \mathbf{n}_\text{s}^\intercal\boldsymbol{\mathcal{K}}
\boldsymbol{\phi}}$&$2e^{2u}$\\
neutral triplet&$e^{\pm \mathbf{n}_\text{t}^\intercal\boldsymbol{\mathcal{K}}
\boldsymbol{\phi}}$&$2e^{-2u}$\\
\hline
\end{tabularx}
\caption{The six types of gap-opening perturbations that are marginal when
$h = v_3=v_4=0$. There are two representatives of each neutral type and four of
each charged type.  Here ${\mathbf{c}_\text{h}^\intercal=(1,1,-1,-1)}$,
$\mathbf{c}_\text{p}^\intercal=(1,1,1,1)$, $\mathbf{n}_\text{s}^\intercal=(1,-1,-1,1)$,
and $\mathbf{n}_\text{t}^\intercal=(1,-1,1,-1)$. The important parameters $u$ and $v$
are defined as
$\tanh{2u}=-\frac{(v_3-v_4)}{v_1-v_2}$ and $\tanh{2v}=-\frac{(v_3+v_4)}{v_1+v_2}$.}
\label{table1}
\end{table}

We now focus our attention on a system in which an ordinary $s$-wave
superconductor couples to the trench, and therefore proximity-induces singlet
pairing. Likewise, we will assume that electron hopping across the trench
acts merely to restore the original quantum Hall state, rather than
introducing additional spin flips. We therefore choose parameters such that
singlet pairing, singlet hopping, and neutral singlet coupling are the
dominant perturbations. Note that these terms are unaffected by the Zeeman
field, while the triplet terms will have oscillating coefficients. Using
Table~\ref{table1}, we can find a region of parameter space for which all
three singlet terms are \emph{simultaneously} relevant by simply setting
$u<0$, $v\approx0$. This is reasonable as a physical regime, given that we
must have $\abs{v_1}>\abs{v_2}$ for stability of the individual edges and
repulsive interactions will generically give $v_2>0$, since $v_2$ is the
density-density interaction between charges on the same edge. Likewise $v_3$
represents repulsion between charges of the same spin on opposite edges, and
$v_4$ encodes repulsion between charges of opposite spins on opposite edges.
It is not unreasonable to expect $v_3>v_4$, and thus $u<0$. We shall work in
this parameter regime for the remainder of this analysis.

The fact that singlet pairing and direct tunneling across the trench can
become simultaneously relevant for the same set of (reasonable) parameters is
a great virtue of the unpolarized $\nu = 2/3$ state.  That is, the
superconductor can induce a gap along the trench \emph{under the same
conditions in which the $2/3$ state itself would also reform}. This feature
is difficult to achieve in other Abelian quantum Hall states.  [Consider,
\emph{e.g.}, the $\nu=2/5$ state with the same interaction matrix in
Eq.~(S\ref{Vdef}).  Here the scaling dimensions for the superconducting and
tunneling perturbations respectively read
$\Delta_S=\frac{1}{2}e^{2u}+\frac{5}{2}e^{2v}$ and
$\Delta_T=\frac{1}{2}e^{2u}+\frac{5}{2}e^{-2v}$. These two terms cannot both
be made relevant (\emph{i.e.} $\Delta_{S,T}<2$) for the same values of $u$
and $v$.]

Crucially, although both singlet pairing and hopping terms are relevant at
$\nu = 2/3$, the gaps favored by these two processes are incompatible with
one another.  Rather, in any given region of the trench either the pairing
\emph{or} the hopping mechanism must win out in order for a gap to open.
This is because the (charge-sector) fields that these two types of coupling
try to pin are dual to one another. If one of them is pinned, the other must
fluctuate. For $\nu=2/3$, though, there is no such issue in the neutral
sector---\emph{i.e.}, all three singlet terms in Table \ref{table1} favor
gapping the neutral fields in compatible ways. The neutral sector thus gaps
out trivially, independent of which mechanism wins out.

As is the case when there is only one type of (fractional) edge mode, the
boundary between a region of the trench gapped by pairing and one gapped by
direct hopping supports a parafermionic zero-energy
mode~\cite{Clarke2013a,Lindner2012,Cheng2012}. In this case, the zero mode
has a $\mathbb{Z}_3$ character.  The main text describes a number of
nontrivial consequences for transport when these (and other types of zero
modes) are probed with quantum Hall edges.  As an aside, we also comment that
the unpolarized $2/3$ case may be of particular interest for more traditional
topological quantum information applications because of the complete gapping
of the neutral sector. This gap partially inoculates the zero modes here
against noise.  In particular, stray electrons \emph{cannot} directly affect
the state of quantum information encoded in these zero modes, because they
possess neither the correct charge ($2e/3$, $4e/3$, or $0 \mod 2e$) nor spin
(always 0) to do so.

\putbib
\end{bibunit}

\begin{thebibliography}{42}
\expandafter\ifx\csname natexlab\endcsname\relax\def\natexlab#1{#1}\fi
\expandafter\ifx\csname bibnamefont\endcsname\relax
  \def\bibnamefont#1{#1}\fi
\expandafter\ifx\csname bibfnamefont\endcsname\relax
  \def\bibfnamefont#1{#1}\fi
\expandafter\ifx\csname citenamefont\endcsname\relax
  \def\citenamefont#1{#1}\fi
\expandafter\ifx\csname url\endcsname\relax
  \def\url#1{\texttt{#1}}\fi
\expandafter\ifx\csname urlprefix\endcsname\relax\def\urlprefix{URL }\fi
\providecommand{\bibinfo}[2]{#2}
\providecommand{\eprint}[2][]{\url{#2}}

\bibitem[{\citenamefont{Anderson}(1972)}]{Anderson1972}
\bibinfo{author}{\bibfnamefont{P.~W.} \bibnamefont{Anderson}},
  \bibinfo{journal}{Science} \textbf{\bibinfo{volume}{177}},
  \bibinfo{pages}{393} (\bibinfo{year}{1972}).

\bibitem[{\citenamefont{Nayak et~al.}(2008)\citenamefont{Nayak, Simon, Stern,
  Freedman, and Das~Sarma}}]{Nayak2008}
\bibinfo{author}{\bibfnamefont{C.}~\bibnamefont{Nayak}},
  \bibinfo{author}{\bibfnamefont{S.~H.} \bibnamefont{Simon}},
  \bibinfo{author}{\bibfnamefont{A.}~\bibnamefont{Stern}},
  \bibinfo{author}{\bibfnamefont{M.}~\bibnamefont{Freedman}}, \bibnamefont{and}
  \bibinfo{author}{\bibfnamefont{S.}~\bibnamefont{Das~Sarma}},
  \bibinfo{journal}{Rev. Mod. Phys.} \textbf{\bibinfo{volume}{80}},
  \bibinfo{eid}{1083} (\bibinfo{year}{2008}), \eprint{arXiv:0707.1889}.

\bibitem[{\citenamefont{Kitaev}(2001)}]{Kitaev2001}
\bibinfo{author}{\bibfnamefont{A.~Y.} \bibnamefont{Kitaev}},
  \bibinfo{journal}{Phys.-Usp.} \textbf{\bibinfo{volume}{44}},
  \bibinfo{pages}{131} (\bibinfo{year}{2001}), \eprint{cond-mat/0010440}.

\bibitem[{\citenamefont{Fu and Kane}(2008)}]{Fu2008}
\bibinfo{author}{\bibfnamefont{L.}~\bibnamefont{Fu}} \bibnamefont{and}
  \bibinfo{author}{\bibfnamefont{C.~L.} \bibnamefont{Kane}},
  \bibinfo{journal}{Phys. Rev. Lett.} \textbf{\bibinfo{volume}{100}},
  \bibinfo{eid}{096407} (\bibinfo{year}{2008}), \eprint{arXiv:0707.1692}.

\bibitem[{\citenamefont{Lutchyn et~al.}(2010)\citenamefont{Lutchyn, Sau, and
  Das~Sarma}}]{Lutchyn2010a}
\bibinfo{author}{\bibfnamefont{R.~M.} \bibnamefont{Lutchyn}},
  \bibinfo{author}{\bibfnamefont{J.~D.} \bibnamefont{Sau}}, \bibnamefont{and}
  \bibinfo{author}{\bibfnamefont{S.}~\bibnamefont{Das~Sarma}},
  \bibinfo{journal}{Phys. Rev. Lett.} \textbf{\bibinfo{volume}{105}},
  \bibinfo{pages}{077001} (\bibinfo{year}{2010}), \eprint{arXiv:1002.4033}.

\bibitem[{\citenamefont{Oreg et~al.}(2010)\citenamefont{Oreg, Refael, and von
  Oppen}}]{Oreg2010}
\bibinfo{author}{\bibfnamefont{Y.}~\bibnamefont{Oreg}},
  \bibinfo{author}{\bibfnamefont{G.}~\bibnamefont{Refael}}, \bibnamefont{and}
  \bibinfo{author}{\bibfnamefont{F.}~\bibnamefont{von Oppen}},
  \bibinfo{journal}{Phys. Rev. Lett.} \textbf{\bibinfo{volume}{105}},
  \bibinfo{pages}{177002} (\bibinfo{year}{2010}), \eprint{arXiv:1003.1145}.

\bibitem[{\citenamefont{Choy et~al.}(2011)\citenamefont{Choy, Edge, Akhmerov,
  and Beenakker}}]{Choy2011}
\bibinfo{author}{\bibfnamefont{T.-P.} \bibnamefont{Choy}},
  \bibinfo{author}{\bibfnamefont{J.~M.} \bibnamefont{Edge}},
  \bibinfo{author}{\bibfnamefont{A.~R.} \bibnamefont{Akhmerov}},
  \bibnamefont{and} \bibinfo{author}{\bibfnamefont{C.~W.~J.}
  \bibnamefont{Beenakker}}, \bibinfo{journal}{Phys. Rev. B}
  \textbf{\bibinfo{volume}{84}}, \bibinfo{pages}{195442}
  (\bibinfo{year}{2011}), \eprint{arXiv:1108.0419}.

\bibitem[{\citenamefont{Nadj-Perge et~al.}(2013)\citenamefont{Nadj-Perge,
  Drozdov, Bernevig, and Yazdani}}]{Nadj-Perge2013}
\bibinfo{author}{\bibfnamefont{S.}~\bibnamefont{Nadj-Perge}},
  \bibinfo{author}{\bibfnamefont{I.~K.} \bibnamefont{Drozdov}},
  \bibinfo{author}{\bibfnamefont{B.~A.} \bibnamefont{Bernevig}},
  \bibnamefont{and} \bibinfo{author}{\bibfnamefont{A.}~\bibnamefont{Yazdani}},
  \bibinfo{journal}{Phys. Rev. B} \textbf{\bibinfo{volume}{88}},
  \bibinfo{pages}{020407} (\bibinfo{year}{2013}), \eprint{arXiv:1303.6363}.

\bibitem[{\citenamefont{Clarke et~al.}(2013)\citenamefont{Clarke, Alicea, and
  Shtengel}}]{Clarke2013a}
\bibinfo{author}{\bibfnamefont{D.~J.} \bibnamefont{Clarke}},
  \bibinfo{author}{\bibfnamefont{J.}~\bibnamefont{Alicea}}, \bibnamefont{and}
  \bibinfo{author}{\bibfnamefont{K.}~\bibnamefont{Shtengel}},
  \bibinfo{journal}{Nat. Commun.} \textbf{\bibinfo{volume}{4}},
  \bibinfo{pages}{1348} (\bibinfo{year}{2013}), \eprint{arXiv:1204.5479}.

\bibitem[{\citenamefont{Lindner et~al.}(2012)\citenamefont{Lindner, Berg,
  Refael, and Stern}}]{Lindner2012}
\bibinfo{author}{\bibfnamefont{N.~H.} \bibnamefont{Lindner}},
  \bibinfo{author}{\bibfnamefont{E.}~\bibnamefont{Berg}},
  \bibinfo{author}{\bibfnamefont{G.}~\bibnamefont{Refael}}, \bibnamefont{and}
  \bibinfo{author}{\bibfnamefont{A.}~\bibnamefont{Stern}},
  \bibinfo{journal}{Phys. Rev. X} \textbf{\bibinfo{volume}{2}},
  \bibinfo{pages}{041002} (\bibinfo{year}{2012}), \eprint{arXiv:1204.5733}.

\bibitem[{\citenamefont{Beenakker}(2013)}]{Beenakker2013a}
\bibinfo{author}{\bibfnamefont{C.~W.~J.} \bibnamefont{Beenakker}},
  \bibinfo{journal}{Annu. Rev. Condens. Matter Phys.}
  \textbf{\bibinfo{volume}{4}}, \bibinfo{pages}{113} (\bibinfo{year}{2013}),
  \eprint{arXiv:1112.1950}.

\bibitem[{\citenamefont{Alicea}(2012)}]{Alicea2012a}
\bibinfo{author}{\bibfnamefont{J.}~\bibnamefont{Alicea}},
  \bibinfo{journal}{Rep. Prog. Phys.} \textbf{\bibinfo{volume}{75}},
  \bibinfo{pages}{076501} (\bibinfo{year}{2012}), \eprint{arXiv:1202.1293}.

\bibitem[{\citenamefont{Leijnse and Flensberg}(2012)}]{Leijnse2012}
\bibinfo{author}{\bibfnamefont{M.}~\bibnamefont{Leijnse}} \bibnamefont{and}
  \bibinfo{author}{\bibfnamefont{K.}~\bibnamefont{Flensberg}},
  \bibinfo{journal}{Semicond. Sci. Technol.} \textbf{\bibinfo{volume}{27}},
  \bibinfo{pages}{124003} (\bibinfo{year}{2012}), \eprint{arXiv:1206.1736}.

\bibitem[{\citenamefont{Stanescu and Tewari}(2013)}]{Stanescu2013a}
\bibinfo{author}{\bibfnamefont{T.~D.} \bibnamefont{Stanescu}} \bibnamefont{and}
  \bibinfo{author}{\bibfnamefont{S.}~\bibnamefont{Tewari}},
  \bibinfo{journal}{J. Phys.: Condens. Matter} \textbf{\bibinfo{volume}{25}},
  \bibinfo{pages}{233201} (\bibinfo{year}{2013}), \eprint{arXiv:1302.5433}.

\bibitem[{\citenamefont{Cheng}(2012)}]{Cheng2012}
\bibinfo{author}{\bibfnamefont{M.}~\bibnamefont{Cheng}},
  \bibinfo{journal}{Phys. Rev. B} \textbf{\bibinfo{volume}{86}},
  \bibinfo{pages}{195126} (\bibinfo{year}{2012}), \eprint{arXiv:1204.6084}.

\bibitem[{\citenamefont{Fendley}(2012)}]{Fendley2012}
\bibinfo{author}{\bibfnamefont{P.}~\bibnamefont{Fendley}}, \bibinfo{journal}{J.
  Stat. Mech.} p. \bibinfo{pages}{P11020} (\bibinfo{year}{2012}),
  \eprint{arXiv:1209.0472}.

\bibitem[{\citenamefont{Barkeshli and Qi}(2012)}]{Barkeshli2012a}
\bibinfo{author}{\bibfnamefont{M.}~\bibnamefont{Barkeshli}} \bibnamefont{and}
  \bibinfo{author}{\bibfnamefont{X.-L.} \bibnamefont{Qi}},
  \bibinfo{journal}{Phys. Rev. X} \textbf{\bibinfo{volume}{2}},
  \bibinfo{pages}{031013} (\bibinfo{year}{2012}), \eprint{arXiv:1112.3311}.

\bibitem[{\citenamefont{Vaezi}(2013)}]{Vaezi2013}
\bibinfo{author}{\bibfnamefont{A.}~\bibnamefont{Vaezi}},
  \bibinfo{journal}{Phys. Rev. B} \textbf{\bibinfo{volume}{87}},
  \bibinfo{pages}{035132} (\bibinfo{year}{2013}), \eprint{arXiv:1204.6245}.

\bibitem[{\citenamefont{Oreg et~al.}(2013)\citenamefont{Oreg, Sela, and
  Stern}}]{Oreg2013}
\bibinfo{author}{\bibfnamefont{Y.}~\bibnamefont{Oreg}},
  \bibinfo{author}{\bibfnamefont{E.}~\bibnamefont{Sela}}, \bibnamefont{and}
  \bibinfo{author}{\bibfnamefont{A.}~\bibnamefont{Stern}}
  (\bibinfo{year}{2013}), \eprint{arXiv:1301.7335}.

\bibitem[{\citenamefont{Klinovaja and
  Loss}(2013{\natexlab{a}})}]{Klinovaja2014a}
\bibinfo{author}{\bibfnamefont{J.}~\bibnamefont{Klinovaja}} \bibnamefont{and}
  \bibinfo{author}{\bibfnamefont{D.}~\bibnamefont{Loss}}
  (\bibinfo{year}{2013}{\natexlab{a}}), \eprint{arXiv:1311.3259}.

\bibitem[{\citenamefont{Klinovaja and
  Loss}(2013{\natexlab{b}})}]{Klinovaja2014b}
\bibinfo{author}{\bibfnamefont{J.}~\bibnamefont{Klinovaja}} \bibnamefont{and}
  \bibinfo{author}{\bibfnamefont{D.}~\bibnamefont{Loss}}
  (\bibinfo{year}{2013}{\natexlab{b}}), \eprint{arXiv:1312.1998}.

\bibitem[{\citenamefont{Barkeshli et~al.}(2013)\citenamefont{Barkeshli, Jian,
  and Qi}}]{Barkeshli2013a}
\bibinfo{author}{\bibfnamefont{M.}~\bibnamefont{Barkeshli}},
  \bibinfo{author}{\bibfnamefont{C.-M.} \bibnamefont{Jian}}, \bibnamefont{and}
  \bibinfo{author}{\bibfnamefont{X.-L.} \bibnamefont{Qi}},
  \bibinfo{journal}{Phys. Rev. B} \textbf{\bibinfo{volume}{87}},
  \bibinfo{pages}{045130} (\bibinfo{year}{2013}), \eprint{arXiv:1208.4834}.

\bibitem[{\citenamefont{Barkeshli and Qi}(2013)}]{Barkeshli2013b}
\bibinfo{author}{\bibfnamefont{M.}~\bibnamefont{Barkeshli}} \bibnamefont{and}
  \bibinfo{author}{\bibfnamefont{X.-L.} \bibnamefont{Qi}}
  (\bibinfo{year}{2013}), \eprint{arXiv:1302.2673}.

\bibitem[{\citenamefont{Sengupta et~al.}(2001)\citenamefont{Sengupta,
  \v{Z}uti\'{c}, Kwon, Yakovenko, and Das~Sarma}}]{Sengupta2001}
\bibinfo{author}{\bibfnamefont{K.}~\bibnamefont{Sengupta}},
  \bibinfo{author}{\bibfnamefont{I.}~\bibnamefont{\v{Z}uti\'{c}}},
  \bibinfo{author}{\bibfnamefont{H.-J.} \bibnamefont{Kwon}},
  \bibinfo{author}{\bibfnamefont{V.~M.} \bibnamefont{Yakovenko}},
  \bibnamefont{and}
  \bibinfo{author}{\bibfnamefont{S.}~\bibnamefont{Das~Sarma}},
  \bibinfo{journal}{Phys. Rev. B} \textbf{\bibinfo{volume}{63}},
  \bibinfo{pages}{144531} (\bibinfo{year}{2001}), \eprint{cond-mat/0010206}.

\bibitem[{\citenamefont{Bolech and Demler}(2007)}]{Bolech2007}
\bibinfo{author}{\bibfnamefont{C.~J.} \bibnamefont{Bolech}} \bibnamefont{and}
  \bibinfo{author}{\bibfnamefont{E.}~\bibnamefont{Demler}},
  \bibinfo{journal}{Phys. Rev. Lett.} \textbf{\bibinfo{volume}{98}},
  \bibinfo{pages}{237002} (\bibinfo{year}{2007}), \eprint{cond-mat/0607779}.

\bibitem[{\citenamefont{Law et~al.}(2009)\citenamefont{Law, Lee, and
  Ng}}]{Law2009}
\bibinfo{author}{\bibfnamefont{K.~T.} \bibnamefont{Law}},
  \bibinfo{author}{\bibfnamefont{P.~A.} \bibnamefont{Lee}}, \bibnamefont{and}
  \bibinfo{author}{\bibfnamefont{T.~K.} \bibnamefont{Ng}},
  \bibinfo{journal}{Phys. Rev. Lett.} \textbf{\bibinfo{volume}{103}},
  \bibinfo{pages}{237001} (\bibinfo{year}{2009}), \eprint{arXiv:0907.1909}.

\bibitem[{\citenamefont{Flensberg}(2010)}]{Flensberg2010}
\bibinfo{author}{\bibfnamefont{K.}~\bibnamefont{Flensberg}},
  \bibinfo{journal}{Phys. Rev. B} \textbf{\bibinfo{volume}{82}},
  \bibinfo{pages}{180516} (\bibinfo{year}{2010}), \eprint{arXiv:1009.3533}.

\bibitem[{\citenamefont{Wimmer et~al.}(2011)\citenamefont{Wimmer, Akhmerov,
  Dahlhaus, and Beenakker}}]{Wimmer2011}
\bibinfo{author}{\bibfnamefont{M.}~\bibnamefont{Wimmer}},
  \bibinfo{author}{\bibfnamefont{A.~R.} \bibnamefont{Akhmerov}},
  \bibinfo{author}{\bibfnamefont{J.~P.} \bibnamefont{Dahlhaus}},
  \bibnamefont{and} \bibinfo{author}{\bibfnamefont{C.~W.~J.}
  \bibnamefont{Beenakker}}, \bibinfo{journal}{New J. Phys.}
  \textbf{\bibinfo{volume}{13}}, \bibinfo{pages}{053016}
  (\bibinfo{year}{2011}), \eprint{arXiv:1101.5795}.

\bibitem[{\citenamefont{Fidkowski et~al.}(2012)\citenamefont{Fidkowski, Alicea,
  Lindner, Lutchyn, and Fisher}}]{Fidkowski2012}
\bibinfo{author}{\bibfnamefont{L.}~\bibnamefont{Fidkowski}},
  \bibinfo{author}{\bibfnamefont{J.}~\bibnamefont{Alicea}},
  \bibinfo{author}{\bibfnamefont{N.~H.} \bibnamefont{Lindner}},
  \bibinfo{author}{\bibfnamefont{R.~M.} \bibnamefont{Lutchyn}},
  \bibnamefont{and} \bibinfo{author}{\bibfnamefont{M.~P.~A.}
  \bibnamefont{Fisher}}, \bibinfo{journal}{Phys. Rev. B}
  \textbf{\bibinfo{volume}{85}}, \bibinfo{pages}{245121}
  (\bibinfo{year}{2012}), \eprint{arXiv:1203.4818}.

\bibitem[{\citenamefont{Lin et~al.}(2012)\citenamefont{Lin, Sau, and
  Das~Sarma}}]{Lin2012}
\bibinfo{author}{\bibfnamefont{C.-H.} \bibnamefont{Lin}},
  \bibinfo{author}{\bibfnamefont{J.~D.} \bibnamefont{Sau}}, \bibnamefont{and}
  \bibinfo{author}{\bibfnamefont{S.}~\bibnamefont{Das~Sarma}},
  \bibinfo{journal}{Phys. Rev. B} \textbf{\bibinfo{volume}{86}},
  \bibinfo{pages}{224511} (\bibinfo{year}{2012}), \eprint{arXiv:1204.3085}.

\bibitem[{\citenamefont{Affleck and Giuliano}(2013)}]{Affleck2013}
\bibinfo{author}{\bibfnamefont{I.}~\bibnamefont{Affleck}} \bibnamefont{and}
  \bibinfo{author}{\bibfnamefont{D.}~\bibnamefont{Giuliano}},
  \bibinfo{journal}{J. Stat. Mech.} \textbf{\bibinfo{volume}{2013}},
  \bibinfo{pages}{P06011} (\bibinfo{year}{2013}), \eprint{arXiv:1305.1888}.

\bibitem[{\citenamefont{Lutchyn and Skrabacz}(2013)}]{Lutchyn2013}
\bibinfo{author}{\bibfnamefont{R.~M.} \bibnamefont{Lutchyn}} \bibnamefont{and}
  \bibinfo{author}{\bibfnamefont{J.~H.} \bibnamefont{Skrabacz}},
  \bibinfo{journal}{Phys. Rev. B} \textbf{\bibinfo{volume}{88}},
  \bibinfo{pages}{024511} (\bibinfo{year}{2013}), \eprint{arXiv:1302.0289}.

\bibitem[{\citenamefont{Fu and Kane}(2009)}]{Fu2009b}
\bibinfo{author}{\bibfnamefont{L.}~\bibnamefont{Fu}} \bibnamefont{and}
  \bibinfo{author}{\bibfnamefont{C.~L.} \bibnamefont{Kane}},
  \bibinfo{journal}{Phys. Rev. Lett.} \textbf{\bibinfo{volume}{102}},
  \bibinfo{eid}{216403} (\bibinfo{year}{2009}), \eprint{arXiv:0903.2427}.

\bibitem[{\citenamefont{Akhmerov et~al.}(2009)\citenamefont{Akhmerov, Nilsson,
  and Beenakker}}]{Akhmerov2009a}
\bibinfo{author}{\bibfnamefont{A.~R.} \bibnamefont{Akhmerov}},
  \bibinfo{author}{\bibfnamefont{J.}~\bibnamefont{Nilsson}}, \bibnamefont{and}
  \bibinfo{author}{\bibfnamefont{C.~W.~J.} \bibnamefont{Beenakker}},
  \bibinfo{journal}{Phys. Rev. Lett.} \textbf{\bibinfo{volume}{102}},
  \bibinfo{pages}{216404} (\bibinfo{year}{2009}), \eprint{arXiv:0903.2196}.

\bibitem[{\citenamefont{Kane and Fisher}(1995)}]{Kane1995}
\bibinfo{author}{\bibfnamefont{C.~L.} \bibnamefont{Kane}} \bibnamefont{and}
  \bibinfo{author}{\bibfnamefont{M.~P.~A.} \bibnamefont{Fisher}},
  \bibinfo{journal}{Phys. Rev. B} \textbf{\bibinfo{volume}{51}},
  \bibinfo{pages}{13449} (\bibinfo{year}{1995}).

\bibitem[{\citenamefont{Wen}(2004)}]{WenBook}
\bibinfo{author}{\bibfnamefont{X.-G.} \bibnamefont{Wen}},
  \emph{\bibinfo{title}{Quantum Field Theory of Many-Body Systems}}, Oxford
  Graduate Texts (\bibinfo{publisher}{Oxford University Press},
  \bibinfo{address}{Oxford}, \bibinfo{year}{2004}).

\bibitem[{\citenamefont{Finck et~al.}(2011)\citenamefont{Finck, Eisenstein,
  Pfeiffer, and West}}]{Finck2011}
\bibinfo{author}{\bibfnamefont{A.~D.~K.} \bibnamefont{Finck}},
  \bibinfo{author}{\bibfnamefont{J.~P.} \bibnamefont{Eisenstein}},
  \bibinfo{author}{\bibfnamefont{L.~N.} \bibnamefont{Pfeiffer}},
  \bibnamefont{and} \bibinfo{author}{\bibfnamefont{K.~W.} \bibnamefont{West}},
  \bibinfo{journal}{Phys. Rev. Lett.} \textbf{\bibinfo{volume}{106}},
  \bibinfo{pages}{236807} (\bibinfo{year}{2011}), \eprint{arXiv:1012.4220}.

\bibitem[{\citenamefont{Mong et~al.}(2013)\citenamefont{Mong, Clarke, Alicea,
  Lindner, Fendley, Nayak, Oreg, Stern, Berg, Shtengel et~al.}}]{Mong2013}
\bibinfo{author}{\bibfnamefont{R.~S.~K.} \bibnamefont{Mong}},
  \bibinfo{author}{\bibfnamefont{D.~J.} \bibnamefont{Clarke}},
  \bibinfo{author}{\bibfnamefont{J.}~\bibnamefont{Alicea}},
  \bibinfo{author}{\bibfnamefont{N.~H.} \bibnamefont{Lindner}},
  \bibinfo{author}{\bibfnamefont{P.}~\bibnamefont{Fendley}},
  \bibinfo{author}{\bibfnamefont{C.}~\bibnamefont{Nayak}},
  \bibinfo{author}{\bibfnamefont{Y.}~\bibnamefont{Oreg}},
  \bibinfo{author}{\bibfnamefont{A.}~\bibnamefont{Stern}},
  \bibinfo{author}{\bibfnamefont{E.}~\bibnamefont{Berg}},
  \bibinfo{author}{\bibfnamefont{K.}~\bibnamefont{Shtengel}},
  \bibnamefont{et~al.} (\bibinfo{year}{2013}), \bibinfo{note}{to appear in
  {Phys. Rev. X}}, \eprint{arXiv:1307.4403}.

\bibitem[{\citenamefont{Halperin et~al.}(2003)\citenamefont{Halperin, Stern,
  and Girvin}}]{Halperin2003}
\bibinfo{author}{\bibfnamefont{B.~I.} \bibnamefont{Halperin}},
  \bibinfo{author}{\bibfnamefont{A.}~\bibnamefont{Stern}}, \bibnamefont{and}
  \bibinfo{author}{\bibfnamefont{S.~M.} \bibnamefont{Girvin}},
  \bibinfo{journal}{Phys. Rev. B} \textbf{\bibinfo{volume}{67}},
  \bibinfo{pages}{235313} (\bibinfo{year}{2003}), \eprint{cond-mat/0301442}.

\bibitem[{\citenamefont{Nilsson et~al.}(2008)\citenamefont{Nilsson, Akhmerov,
  and Beenakker}}]{Nilsson2008}
\bibinfo{author}{\bibfnamefont{J.}~\bibnamefont{Nilsson}},
  \bibinfo{author}{\bibfnamefont{A.~R.} \bibnamefont{Akhmerov}},
  \bibnamefont{and} \bibinfo{author}{\bibfnamefont{C.~W.~J.}
  \bibnamefont{Beenakker}}, \bibinfo{journal}{Phys. Rev. Lett.}
  \textbf{\bibinfo{volume}{101}}, \bibinfo{pages}{120403}
  (\bibinfo{year}{2008}).

\bibitem[{\citenamefont{Herrmann et~al.}(2010)\citenamefont{Herrmann, Portier,
  Roche, Yeyati, Kontos, and Strunk}}]{Herrmann2010}
\bibinfo{author}{\bibfnamefont{L.~G.} \bibnamefont{Herrmann}},
  \bibinfo{author}{\bibfnamefont{F.}~\bibnamefont{Portier}},
  \bibinfo{author}{\bibfnamefont{P.}~\bibnamefont{Roche}},
  \bibinfo{author}{\bibfnamefont{A.~L.} \bibnamefont{Yeyati}},
  \bibinfo{author}{\bibfnamefont{T.}~\bibnamefont{Kontos}}, \bibnamefont{and}
  \bibinfo{author}{\bibfnamefont{C.}~\bibnamefont{Strunk}},
  \bibinfo{journal}{Phys. Rev. Lett.} \textbf{\bibinfo{volume}{104}},
  \bibinfo{pages}{026801} (\bibinfo{year}{2010}), \eprint{arXiv:0909.3243}.

\bibitem[{\citenamefont{Das et~al.}(2012)\citenamefont{Das, Ronen, Heiblum,
  Mahalu, Kretinin, and Shtrikman}}]{Das2012b}
\bibinfo{author}{\bibfnamefont{A.}~\bibnamefont{Das}},
  \bibinfo{author}{\bibfnamefont{Y.}~\bibnamefont{Ronen}},
  \bibinfo{author}{\bibfnamefont{M.}~\bibnamefont{Heiblum}},
  \bibinfo{author}{\bibfnamefont{D.}~\bibnamefont{Mahalu}},
  \bibinfo{author}{\bibfnamefont{A.~V.} \bibnamefont{Kretinin}},
  \bibnamefont{and}
  \bibinfo{author}{\bibfnamefont{H.}~\bibnamefont{Shtrikman}},
  \bibinfo{journal}{Nat. Commun.} \textbf{\bibinfo{volume}{3}},
  \bibinfo{pages}{1165} (\bibinfo{year}{2012}).

\end{thebibliography}


\begin{thebibliography}{5}
\expandafter\ifx\csname natexlab\endcsname\relax\def\natexlab#1{#1}\fi
\expandafter\ifx\csname bibnamefont\endcsname\relax
  \def\bibnamefont#1{#1}\fi
\expandafter\ifx\csname bibfnamefont\endcsname\relax
  \def\bibfnamefont#1{#1}\fi
\expandafter\ifx\csname citenamefont\endcsname\relax
  \def\citenamefont#1{#1}\fi
\expandafter\ifx\csname url\endcsname\relax
  \def\url#1{\texttt{#1}}\fi
\expandafter\ifx\csname urlprefix\endcsname\relax\def\urlprefix{URL }\fi
\providecommand{\bibinfo}[2]{#2}
\providecommand{\eprint}[2][]{\url{#2}}

\bibitem[{\citenamefont{Wen and Zee}(1992)}]{Wen1992a}
\bibinfo{author}{\bibfnamefont{X.~G.} \bibnamefont{Wen}} \bibnamefont{and}
  \bibinfo{author}{\bibfnamefont{A.}~\bibnamefont{Zee}},
  \bibinfo{journal}{Phys. Rev. B} \textbf{\bibinfo{volume}{46}},
  \bibinfo{pages}{2290} (\bibinfo{year}{1992}).

\bibitem[{\citenamefont{Kane et~al.}(1994)\citenamefont{Kane, Fisher, and
  Polchinski}}]{Kane1994b}
\bibinfo{author}{\bibfnamefont{C.~L.} \bibnamefont{Kane}},
  \bibinfo{author}{\bibfnamefont{M.~P.~A.} \bibnamefont{Fisher}},
  \bibnamefont{and}
  \bibinfo{author}{\bibfnamefont{J.}~\bibnamefont{Polchinski}},
  \bibinfo{journal}{Phys. Rev. Lett.} \textbf{\bibinfo{volume}{72}},
  \bibinfo{pages}{4129} (\bibinfo{year}{1994}), \eprint{cond-mat/9402108}.

\bibitem[{\citenamefont{Clarke et~al.}(2013)\citenamefont{Clarke, Alicea, and
  Shtengel}}]{Clarke2013a}
\bibinfo{author}{\bibfnamefont{D.~J.} \bibnamefont{Clarke}},
  \bibinfo{author}{\bibfnamefont{J.}~\bibnamefont{Alicea}}, \bibnamefont{and}
  \bibinfo{author}{\bibfnamefont{K.}~\bibnamefont{Shtengel}},
  \bibinfo{journal}{Nat. Commun.} \textbf{\bibinfo{volume}{4}},
  \bibinfo{pages}{1348} (\bibinfo{year}{2013}), \eprint{arXiv:1204.5479}.

\bibitem[{\citenamefont{Lindner et~al.}(2012)\citenamefont{Lindner, Berg,
  Refael, and Stern}}]{Lindner2012}
\bibinfo{author}{\bibfnamefont{N.~H.} \bibnamefont{Lindner}},
  \bibinfo{author}{\bibfnamefont{E.}~\bibnamefont{Berg}},
  \bibinfo{author}{\bibfnamefont{G.}~\bibnamefont{Refael}}, \bibnamefont{and}
  \bibinfo{author}{\bibfnamefont{A.}~\bibnamefont{Stern}},
  \bibinfo{journal}{Phys. Rev. X} \textbf{\bibinfo{volume}{2}},
  \bibinfo{pages}{041002} (\bibinfo{year}{2012}), \eprint{arXiv:1204.5733}.

\bibitem[{\citenamefont{Cheng}(2012)}]{Cheng2012}
\bibinfo{author}{\bibfnamefont{M.}~\bibnamefont{Cheng}},
  \bibinfo{journal}{Phys. Rev. B} \textbf{\bibinfo{volume}{86}},
  \bibinfo{pages}{195126} (\bibinfo{year}{2012}), \eprint{arXiv:1204.6084}.

\end{thebibliography}
\end{document}